\documentclass[preprint,superscriptaddress,nofootinbib,showkeys,showpacs,tightenline]{revtex4}  
\usepackage{graphicx}
\usepackage{dcolumn}
\newcommand{\be}{\begin{equation}}
\newcommand{\ee}{\end{equation}}
\newcommand{\bee}{\begin{eqnarray}}
\newcommand{\eee}{\end{eqnarray}}
\newcommand{\eq}{\end{quote}}
\newcommand{\nn}{\nonumber}
\newcommand{\Slash}[1]{\ooalign{\hfil/\hfil\crcr$#1$}}

\def\gsim{\displaystyle\mathop{>}_{\sim}}

\begin{document}      
\preprint{PNU-NTG-3/2005}
\title{Photoproduction of $\Lambda(1520,3/2^{-})$}
\author{Seung-Il Nam}
\email{sinam@rcnp.osaka-u.ac.jp}
\affiliation{Research Center for Nuclear Physics (RCNP), Osaka
  University, Ibaraki, Osaka
567-0047, Japan}
\affiliation{Department of physics and Nuclear physics \& Radiation technology Institute (NuRI),
Pusan University, Keum-jung~gu, Busan 609-735, Korea} 
\author{Atsushi Hosaka}
\email{hosaka@rcnp.osaka-u.ac.jp}
\affiliation{Research Center for Nuclear Physics (RCNP), Osaka
  University, Ibaraki, Osaka
567-0047, Japan}
\author{Hyun-Chul Kim}
\email{hchkim@pusan.ac.kr}
\affiliation{Department of physics and Nuclear physics \& Radiation technology Institute (NuRI),
Pusan University, Keum-jung~gu, Busan 609-735, Korea} 
\begin{abstract}
We investigate the photoproduction of $\Lambda(1520,3/2^-)$ via the 
$\gamma N\to K\Lambda^*$ reaction using the Born approximation and the
Rariata-Schwinger vector-spinor field for the 
spin 3/2 particle.  We reproduce the experimental data of the total 
cross section and the angular dependence for the proton target
qualitatively well and estimate them for the neutron target.  For the
neutron target, we find much smaller total cross section than that of
the proton and the significant $K^*$--exchange contribution.
\end{abstract}
\pacs{13.75.Cs, 14.20.-c}
\keywords{$\Lambda(1520)$, spin 3/2, photoproduction}
\maketitle
\section{introduction}
After the observation of the pentaquark baryon
$\Theta^+(1540)$~\cite{nakano}, the $\Theta^+$ baryon has been
one of the most interesting issues among the recent studies for the
hadron systems.  Taking into account the fact that the LEPS 
collaboration found that $\Lambda(1520)$ is produced
simultaneously in the $\Theta^+$ photoproduction on the deuteron, 
it is of great importance to study on the producing mechanisim of the
$\Lambda(1520)$ and of $\Theta^+$ on the same footing.  Therefore,
in the present report, we present a recent investigation on the
photoproduction of $\Lambda^*\,(\equiv\Lambda(1520))$ via $\gamma N\to
K\Lambda^*$ using the Born approximation. In order to consider the
spin 3/2 baryon relativistically, we employ the Rarita-Shwinger
vector-spinor field formalism~\cite{rarita}. Though the theoretical
calculation contains several model parameters, we observe that the
model-parameter dependences are rather weak in the low energy region, 
where the Born approximation works appropriately. We also find that
the contact term contribution is dominant over other channels.

As for the $\Lambda^*$ photoproduction on the proton target, we can
reproduce qualitatively well the experimental data~\cite{daresbury},
the total cross section and the angular dependence, using the four
dimensional gauge invariant formfactor~\cite{davidson}.  We also
observe that the angular distribution demonstrates a 
strong enhancement in forward scattering, which agrees well with the
experimental data. 

In the case of the neutron target, the order of magnitude of the
total cross section is much smaller than that of the proton one, since
the contact term does not exist in the neutron case.  However, in
this case, being different from the proton case, $K^*$--exchange
contributes much to the angular distribution which shows a peak around
$45^{\circ}$. 

We will organize the present report as follows. In 
Section II we briefly introduce the formalism used in the present
report. We will demonstrate the numerical results in Section III.
Finally we summarize the present work.
\section{General formalism}
We start with the relevant Lagrangians for the
interaction vertices in the $\gamma N\to K\Lambda^*$ reaction.
\bee
\small
\mathcal{L}_{\gamma NN}
&=&-e\bar{p}\left(\gamma_{\mu}+i\frac{\kappa_{p}}{2M_{p}}\sigma_{\mu\nu}k^{\nu}_{1}\right)
A^{\mu}N+{\rm
h.c.},\nn\\\mathcal{L}_{\gamma KK}&=&
ie\left\{ 
(\partial^{\mu}K^{\dagger})K-(\partial^{\mu}K)K^{\dagger}
\right\}A_{\mu}+{\rm
h.c.},\nn\\
\mathcal{L}_{\gamma
\Lambda^*\Lambda^*}&=&-\bar{\Lambda^*}^{\mu}
\left\{\left(-F_{1}\Slash{\epsilon}g_{\mu\nu}+F_3\Slash{\epsilon}\frac{k_{1
\mu}k_{1
\nu}}{2M^{2}_{\Lambda^*}}\right)-\frac{\Slash{k}_{1}\Slash{\epsilon}}
{2M_{\Lambda^*}}\left(-F_{2}g_{\mu\nu}+F_4\frac{k_{1\mu}k_{1 \nu}}
{2M^{2}_{\Lambda^*}}\right)\right\}\Lambda^{*\nu}\,+{\rm 
h.c.},\nn\\
\mathcal{L}_{\gamma
  KK^{*}}&=&g_{\gamma
  KK^{*}}\epsilon_{\mu\nu\sigma\rho}(\partial^{\mu}A^{\nu})
(\partial^{\sigma}K)K^{*\rho}\,+{\rm 
h.c.},\nn\\\mathcal{L}_{KN\Lambda^*}&=&\frac{g_{KN\Lambda^*}}{M_{K}}\bar{\Lambda^*}^{\mu}
\Theta_{\mu\nu}(A,Z)(\partial^{\nu}K){\gamma}_{5}p\,+{\rm 
h.c.},\nn\\
\mathcal{L}_{K^{*}N\Lambda^*}&=&-\frac{ig_{K^{*}N\Lambda^*}}
{M_{V}}\bar{\Lambda^*}^{\mu}\gamma^{\nu}(\partial_{\mu}
K^{*}_{\nu}-\partial_{\nu}K^{*}_{\mu})p+{\rm 
h.c.},\nn\\\mathcal{L}_{\gamma
KN\Lambda^*}&=&-i\frac{eg_{KN\Lambda^*}}{M_{K}}\bar{\Lambda^*}^{\mu}A_{\mu}K{\gamma}_{5}N\,+{\rm
h.c.}. 
\label{Lagrangian}
\normalsize
\eee
Here, $N$, $\Lambda^*_{\mu}$, $K$ and $A^{\mu}$ are the nucleon, $J=3/2^{-}$
Rarita-Schwinger spinor for $\Lambda^*$, pseudo-scalar kaon and photon
fields, respectively.  In order to maintain the
gauge invariance for the present reaction, the contact term
(Kroll-Ruderman term) is necessary.  The interaction for the
$K^{*}N\Lambda^*$ vertex is taken from Ref.~\cite{Machleidt:1987hj}.
As for the $\gamma \Lambda^* \Lambda^*$ vertex 
in the $u$--channel, we utilize the effective interaction suggested
by Ref.~\cite{gourdin} which contains four terms related to the
electric and magnetic multipoles.  We will ignore the electric
coupling $F_1$, since the $\Lambda^*$ is neutral. We will also neglect
$F_3$ and $F_4$ terms, assuming that higher multipole terms are less
important.  Hence, for the photon coupling to $\Lambda^*$, we consider
only the magnetic coupling term $F_2$ whose strength is proportional to
the anomalous magnetic moment of $\Lambda^*$, $\kappa_{\Lambda^*}$,
will be treated as a free parameter. $\Theta_{\mu\nu}(A,Z)$ containing
the off-shell contribution of the spin-3/2 particle is defined as
$\Theta_{\mu\nu}(X)=g_{\mu\nu}+X\gamma_{\mu}\gamma_{\nu}$ in which $X$
is the reduced off-shell parameter~\cite{Read:ye} and will be also
treated as a free parameter. 

In order to determine the coupling constant $g_{KN\Lambda^*}$ we make use of  
the decay width $\Gamma_{\Lambda^*\to KN} = 15.6$ MeV 
with decay ratio 0.45~\cite{Eidelman:2004wy} . We obtain
$g_{KN\Lambda^*}=11.075$.  As for the $K^*N\Lambda^*$ coupling
constant, we will choose the values of $0$, $\pm g_{KN\Lambda^*}$ and
$\pm 2g_{KN\Lambda^*}$ for the numerical calculation, since we have no 
information on how to determine the coupling constant.  The coupling constant
of $g_{\gamma K^{*}K}$ is taken to be $0.254/{\rm GeV}$ for the
charged decay and $0.388/{\rm GeV}$ for the neutral
decay~\cite{Eidelman:2004wy}.  As suggested in Ref.~\cite{davidson}, we
adopt the following parameterization 
for the four dimensional form factors satisfying the crossing symmetry:
\bee
F_{x}(q^2)&=&\frac{\Lambda^4}{\Lambda^4+(x-M^2_x)^2},\,\,x=s,t,u,v\nn\\
F_{c}&=&F_{s}+F_{t}-F_{s}F_{t}. 
\label{formfactor1}
\eee

The form of $F_{c}$ is chosen in such a way that the on-shell values of the
coupling constants are reproduced. 
\section{Numerical results}
First, we present the numerical results of the total cross sections
for the proton target using several different model parameter sets in
Fig.~1. In the figures, we change the model parameters for
$-1<\kappa_{\Lambda^*}<1$, $-1<X<1$ and 
$-22.14<g_{K^*N\Lambda^*}<22.14$ in order to see the dependence on the
model parameters.  The experimental data is taken from 
Ref.~\cite{daresbury}. Though the curves demonstrate sizable model
parameter dependences in the higher energy region but the dependences
become rather weak in the low energy region, where the Born
approximation is known to work well. We note that the total cross
section for the proton target shows the $S$--wave threshold behavior
($\sim [E_{\rm th}-E_{\gamma}]^{1/2}$) due to the contact term
contribution.  Within the several parameter sets, we choose  
$\kappa_{\Lambda^*}=-0.5$, $X=0$ and $g_{K^*N\Lambda^*}=0,\pm
g_{KN\Lambda^*}$ which reproduce the data relatively well for the
numerical calculations.

In Fig.~2,  we show the $t$-dependence for the chosen
parameter sets.  The experimental data are also taken from
Ref.~\cite{daresbury}.  The experimental values are averaged ones for
the incident photon lab energy range 
from $2.8$ GeV to $4.8$ GeV.  We plot the momentum-transfer dependence
for energies in the COM.  For $E_{\rm CM}=$ 2.4 GeV, 2.7 GeV
and 3.0 GeV which lie in the incident photon lab energy range, we can
reproduce the data qualitatively well. We observe that the numerical
results for the momentum transfer dependence is not sensitive to the
model parameters and indicates the strong enhancement in the forward
scattering.   The backward scattering is suppressed due to the four
dimensional formfactor $F_c$ in Eq.~(\ref{formfactor1}).    

Turning to the neutron target case in Fig.~3, we show the
total cross sections with the model parameters which are determined in
the proton target case. We observe that results increase drastically
beyond $E_{\gamma lab}\gsim 3.0$ GeV.  We confirm that this unnatural
behavior comes from the $u$--channel contribution which
contains the model parameters $\kappa_{\Lambda^*}$ and $X$. The
threshold behavior is strongly dependent on the inclusion of the vector
$K^*$--exchange contribution.  Since the contact term, which is the
largest contribution among the kinematical channels, does not exist in
the neutron target case, the order of magnitude of the total cross
section is much smaller than that of the proton case. 

Finally, In Fig.~4 we demonstrate the angular
dependences on the parameter sets. The peaks at $\sim 45^{\circ}$ in
these figures indicate that the $K^*$--exchange contributes
significantly to the $\Lambda^*$ photoproduction reaction on the
neutron target.  This $K^*$--exchange dominant behavior was also shown 
in the $\gamma p\to \bar{K}^0\Theta^+$ reaction, in which the
$K^*$--exchange is the only $t$--channel contribution~\cite{nam3}. 
\section{Summary}
We have studied the $\Lambda(1520)$ photoproduction in the Born
approximation.  The physical oservables, the total cross sections, and
the angular dependence were calculated with appropriate model
parameters used.  The model parameter dependences were not so strong
in the low energy region.  We were able to reproduce the experimental
data qualitatively well for the proton 
target and estimate the total and differential cross sections for the
neutron one which has not been measured yet. We observed
that the contact term contribution is the 
largest one among the kinematical channels.  Therefore, the absence of
the contact term in the neutron target case makes the total cross
section much smaller than that of the proton.  We also note 
$K^*$--exchange plays a significant role in the case of the neutron
target.  The more detailed work on the photoproduction of $\Lambda^*$
is under way~\cite{nam7}.    
\section*{Acknowledgement}
We thank T.~Nakano and A.~Titov for fruitful discussions and
comments. The work of S.I.N. has been supported by
the scholarship endowed from the Ministry of Education, Science,
Sports, and Culture of Japan. The work of H.C.K. is supported by the Korean Research
Foundation (KRF-2003-070-C00015).


\begin{figure}[tbh]
\begin{tabular}{ccc}
\resizebox{5cm}{5cm}{\includegraphics{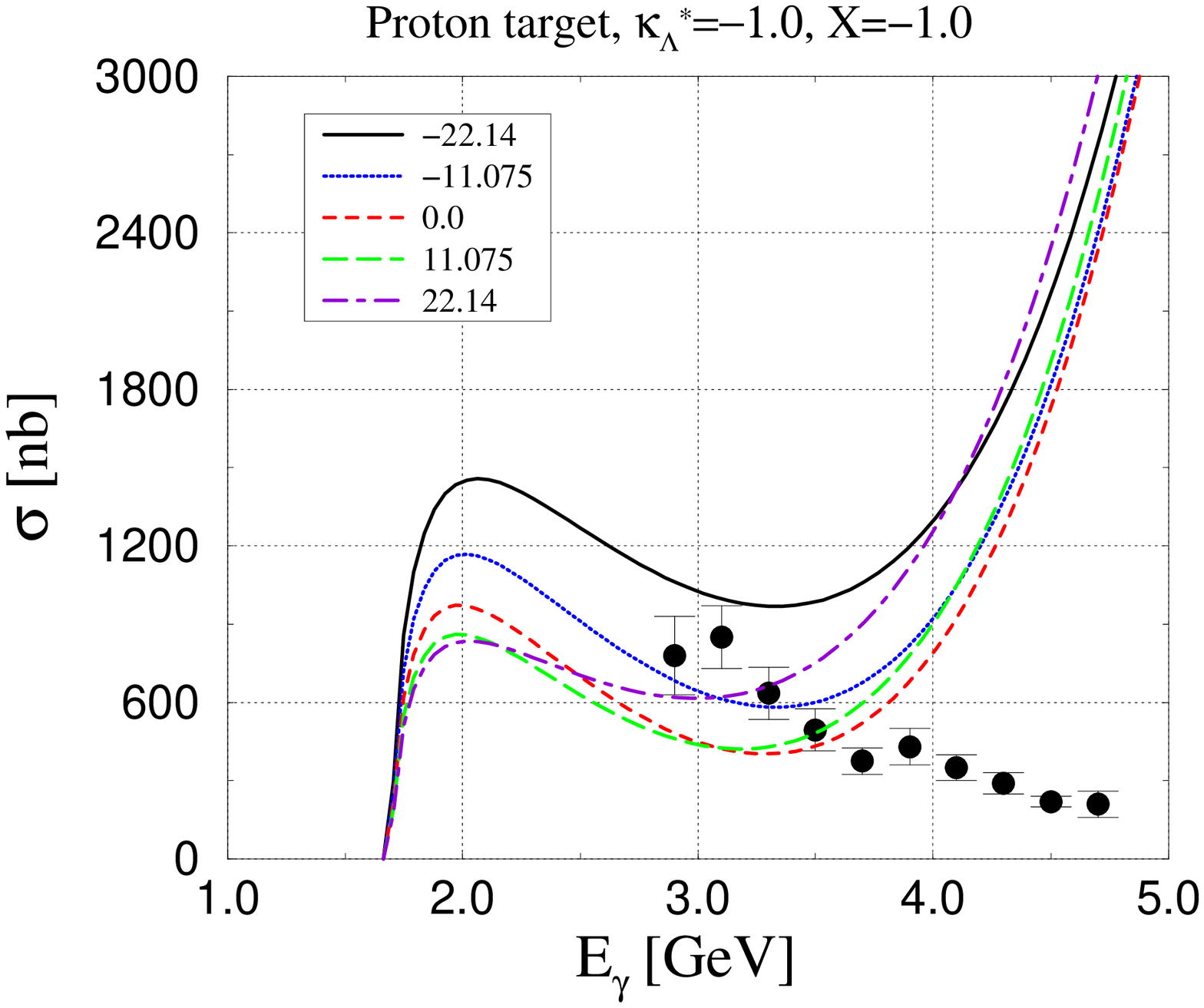}}
\resizebox{5cm}{5cm}{\includegraphics{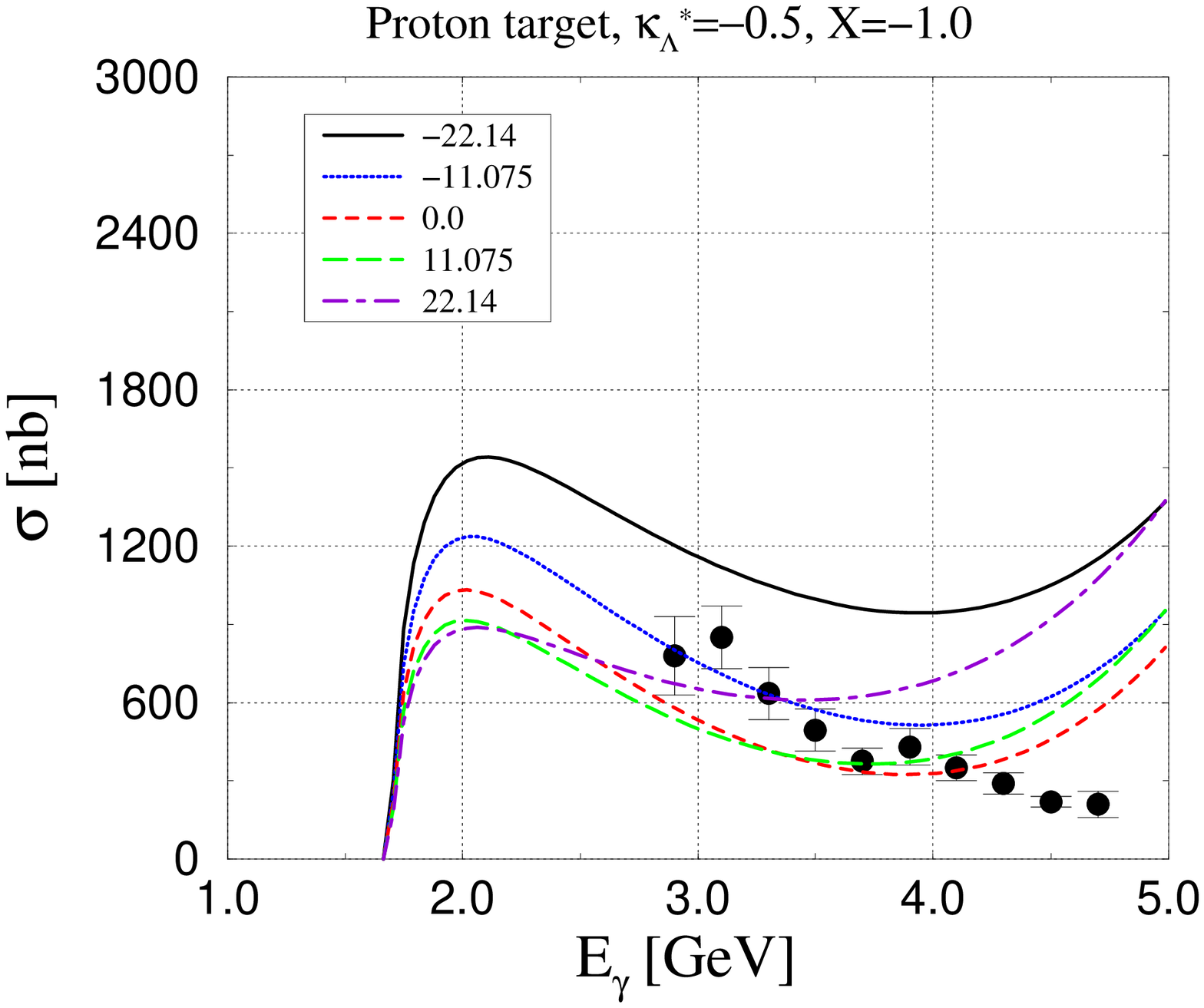}}
\resizebox{5cm}{5cm}{\includegraphics{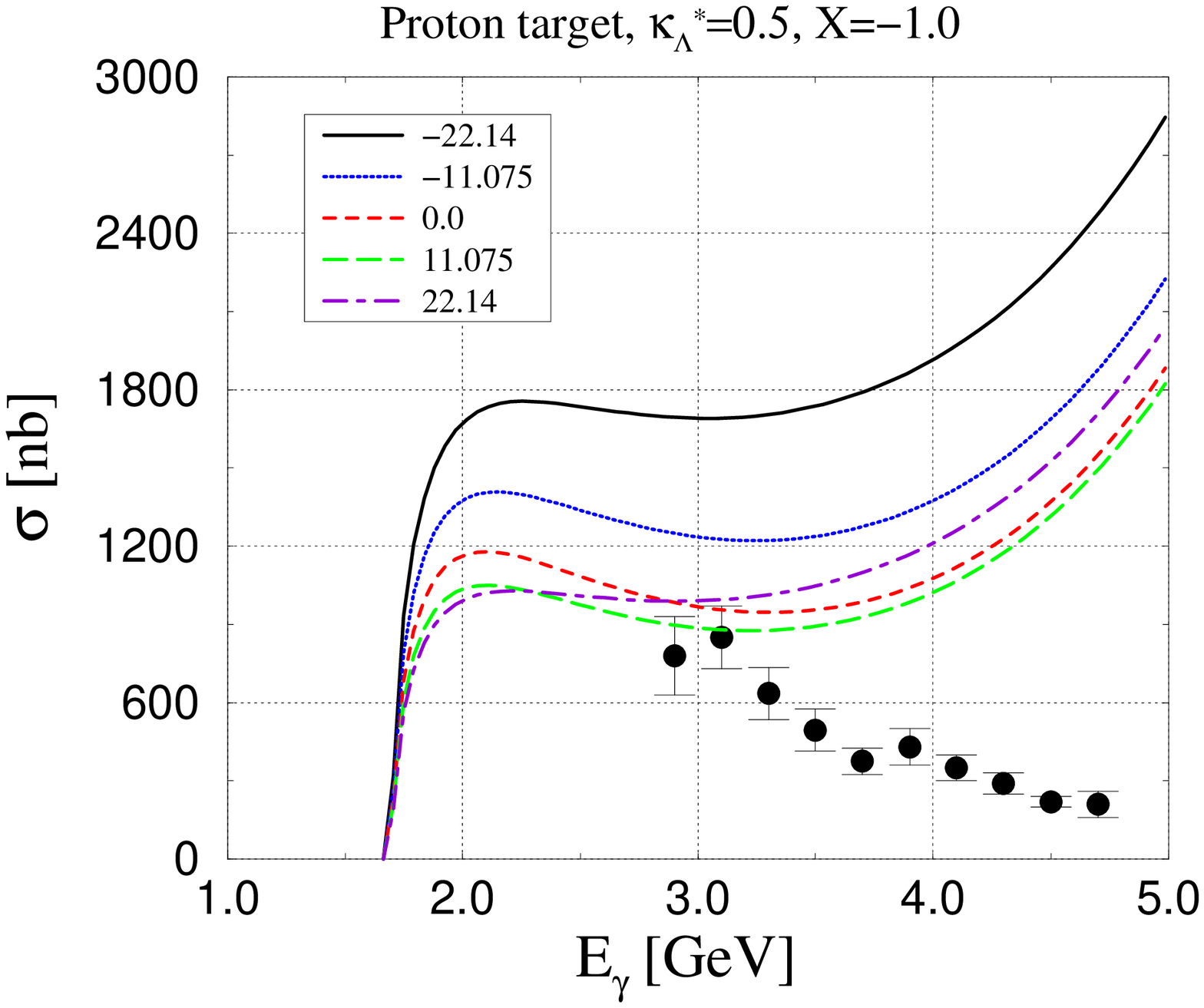}}
\end{tabular}
\begin{tabular}{ccc}
\resizebox{5cm}{5cm}{\includegraphics{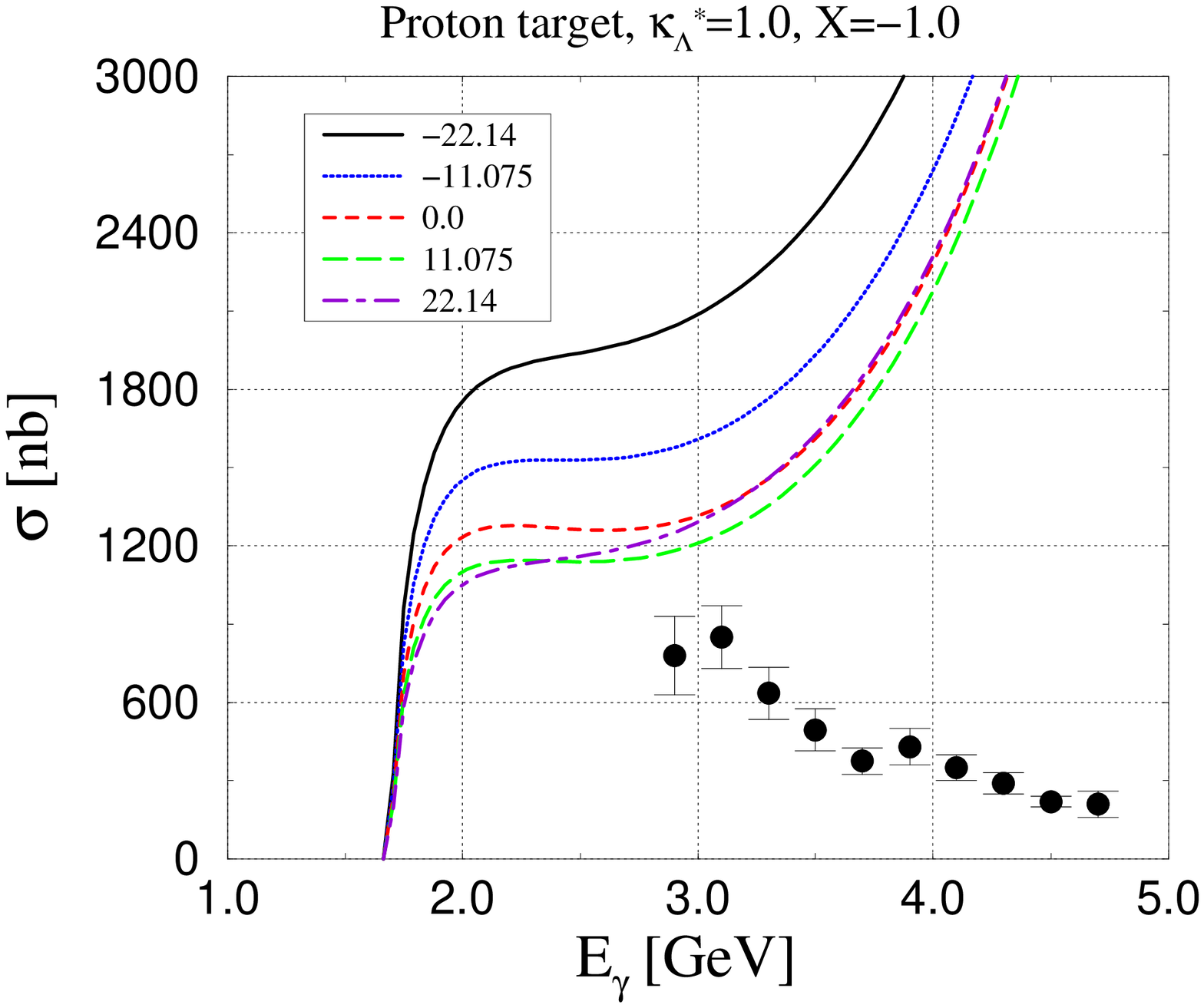}}
\resizebox{5cm}{5cm}{\includegraphics{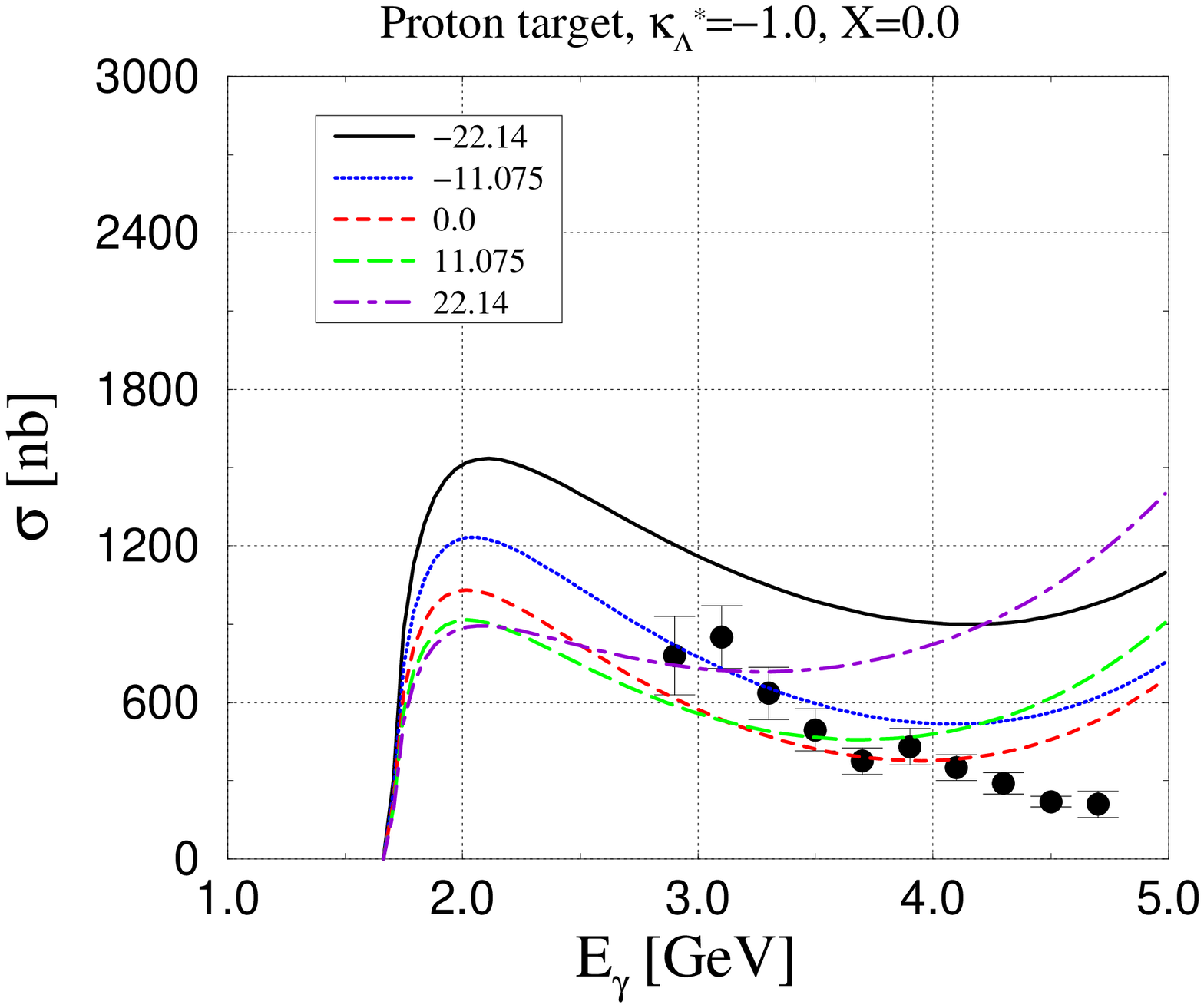}}
\resizebox{5cm}{5cm}{\includegraphics{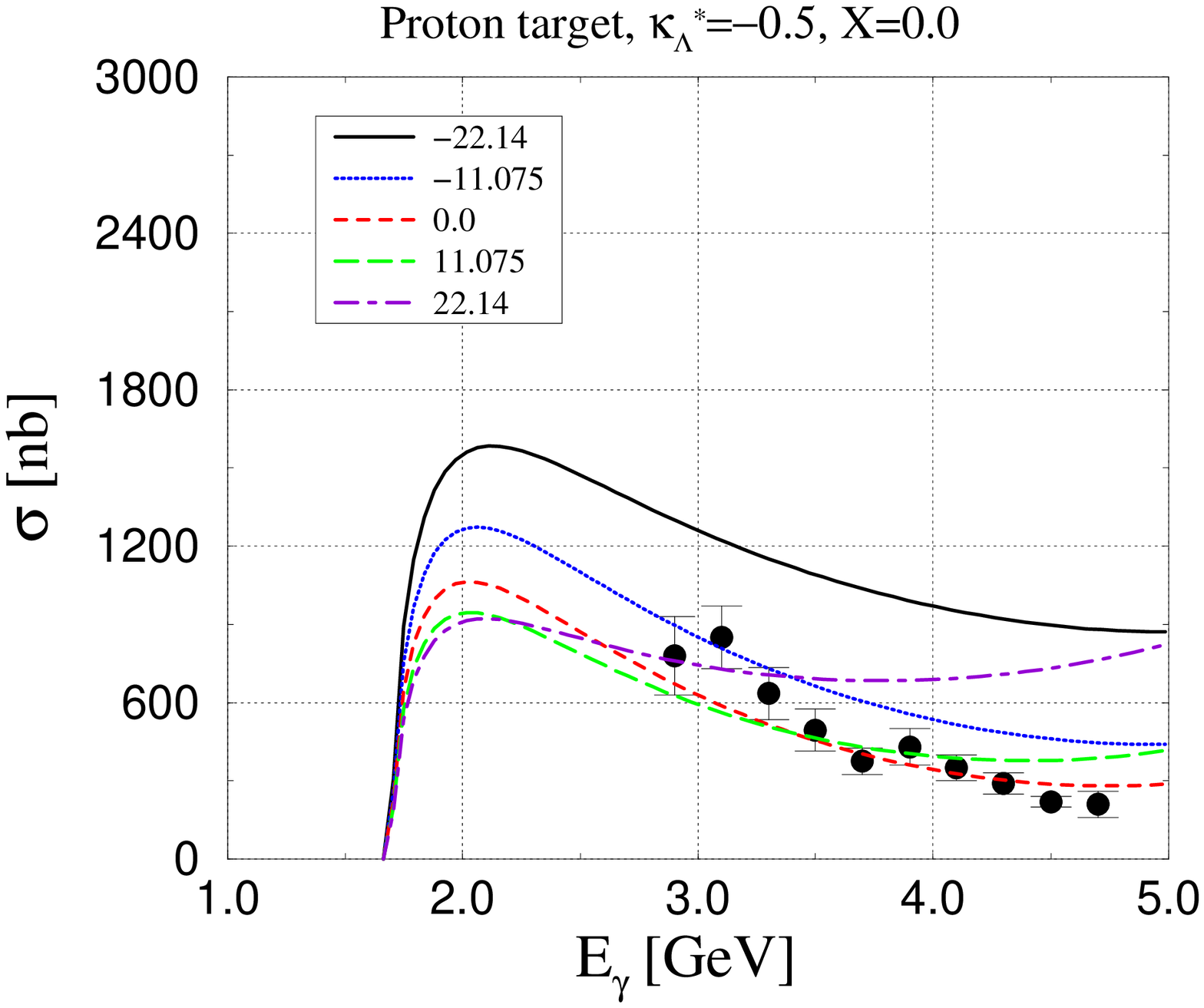}}
\end{tabular}
\begin{tabular}{ccc}
\resizebox{5cm}{5cm}{\includegraphics{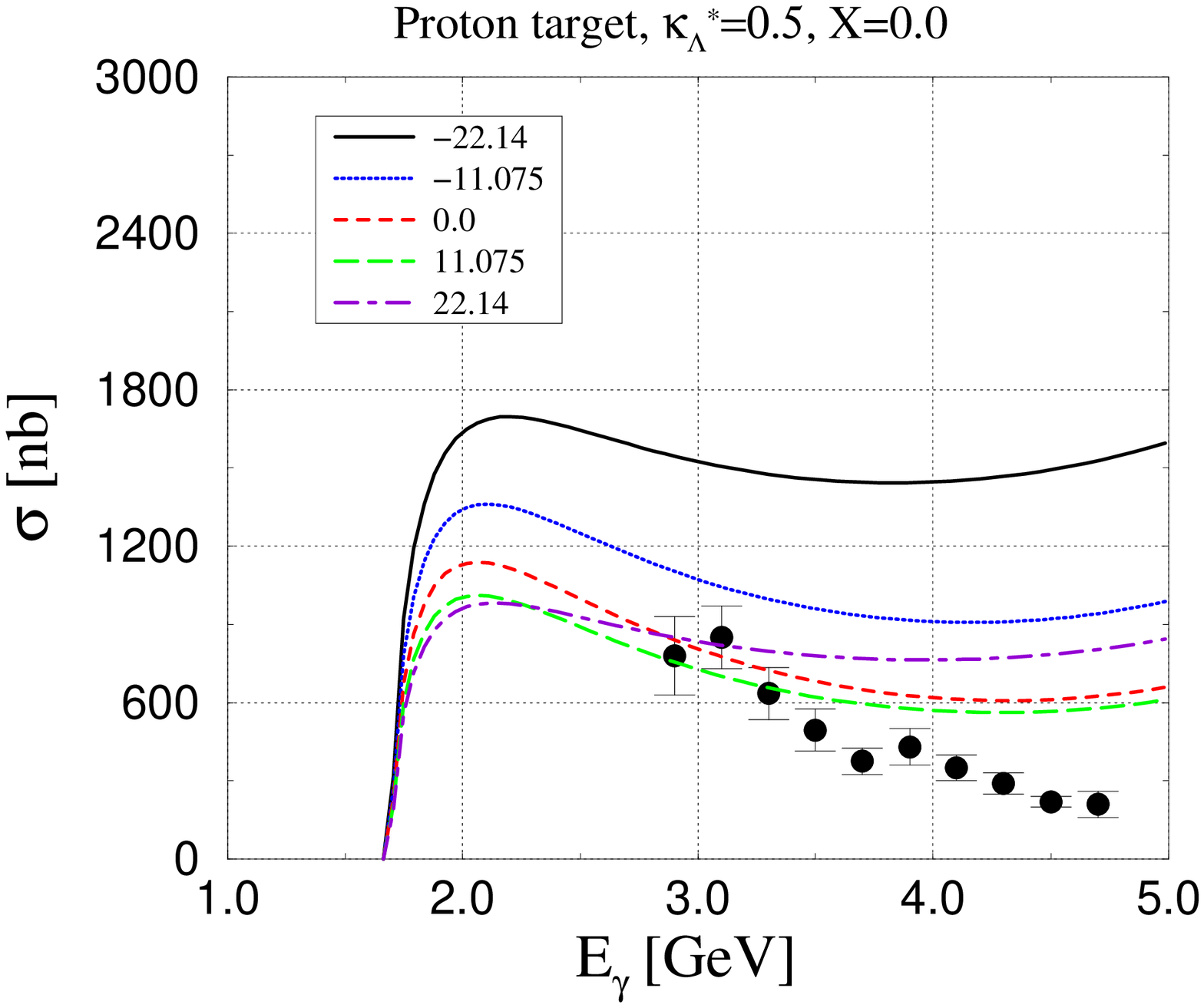}}
\resizebox{5cm}{5cm}{\includegraphics{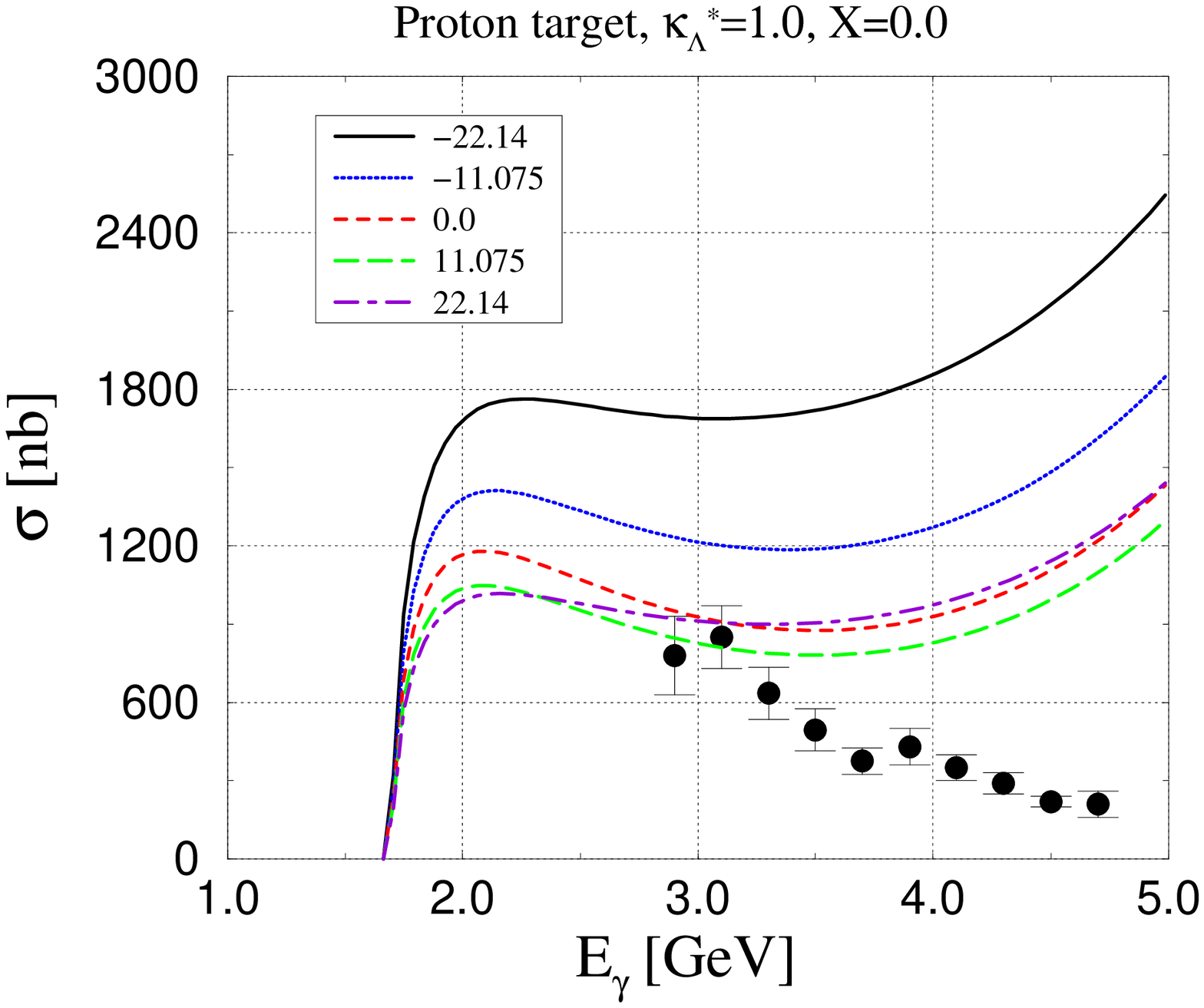}}
\resizebox{5cm}{5cm}{\includegraphics{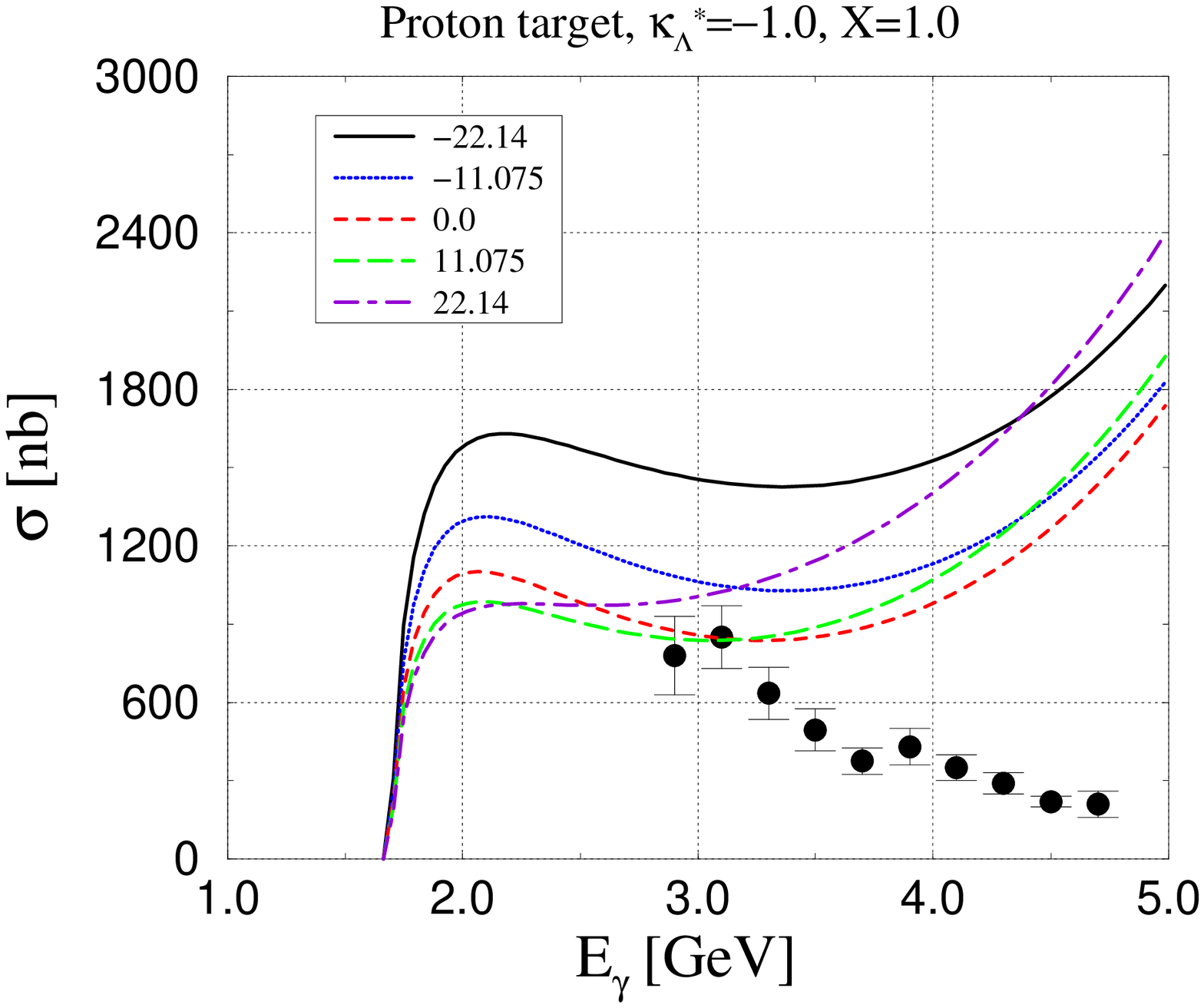}}
\end{tabular}
\begin{tabular}{ccc}
\resizebox{5cm}{5cm}{\includegraphics{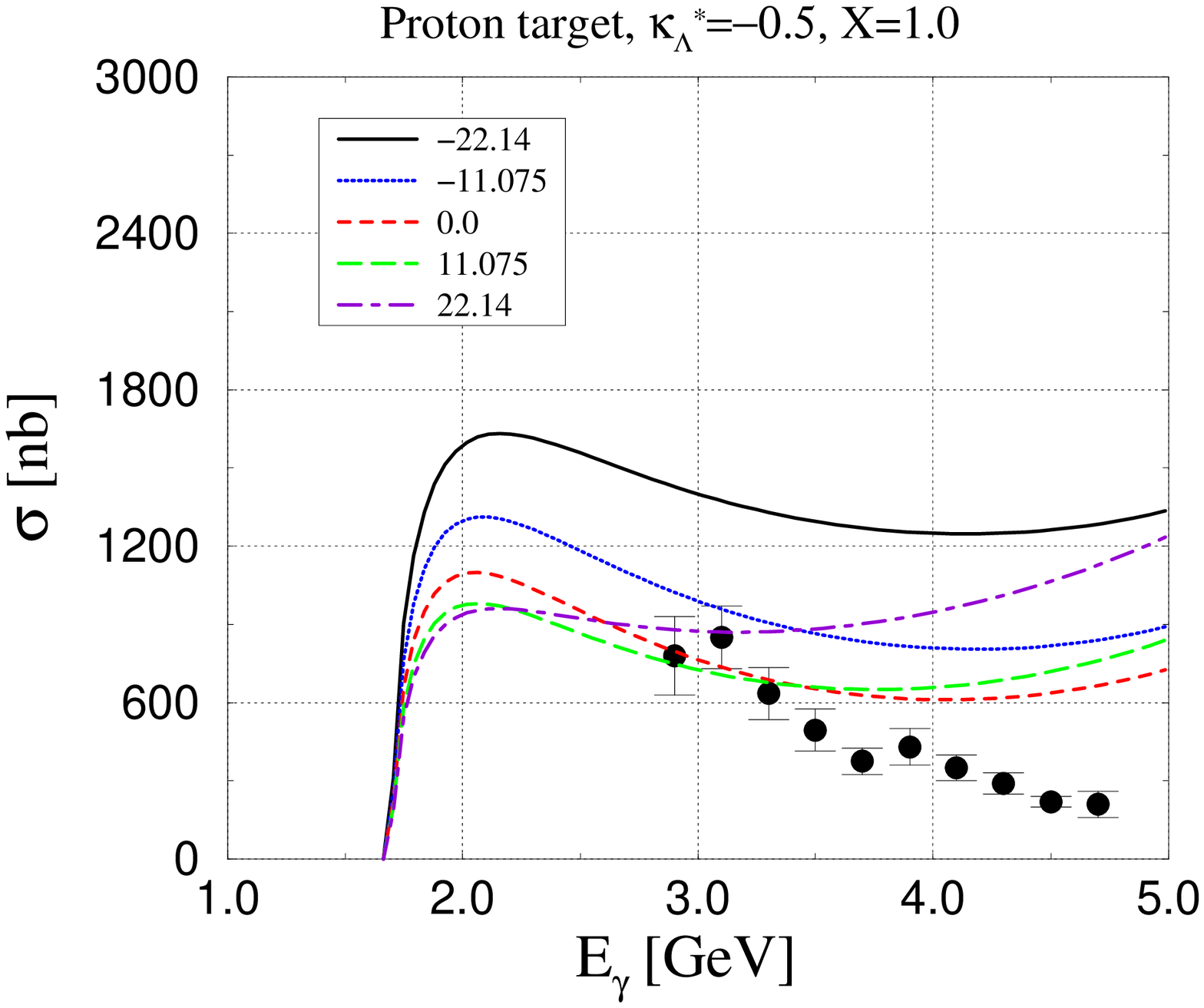}}
\resizebox{5cm}{5cm}{\includegraphics{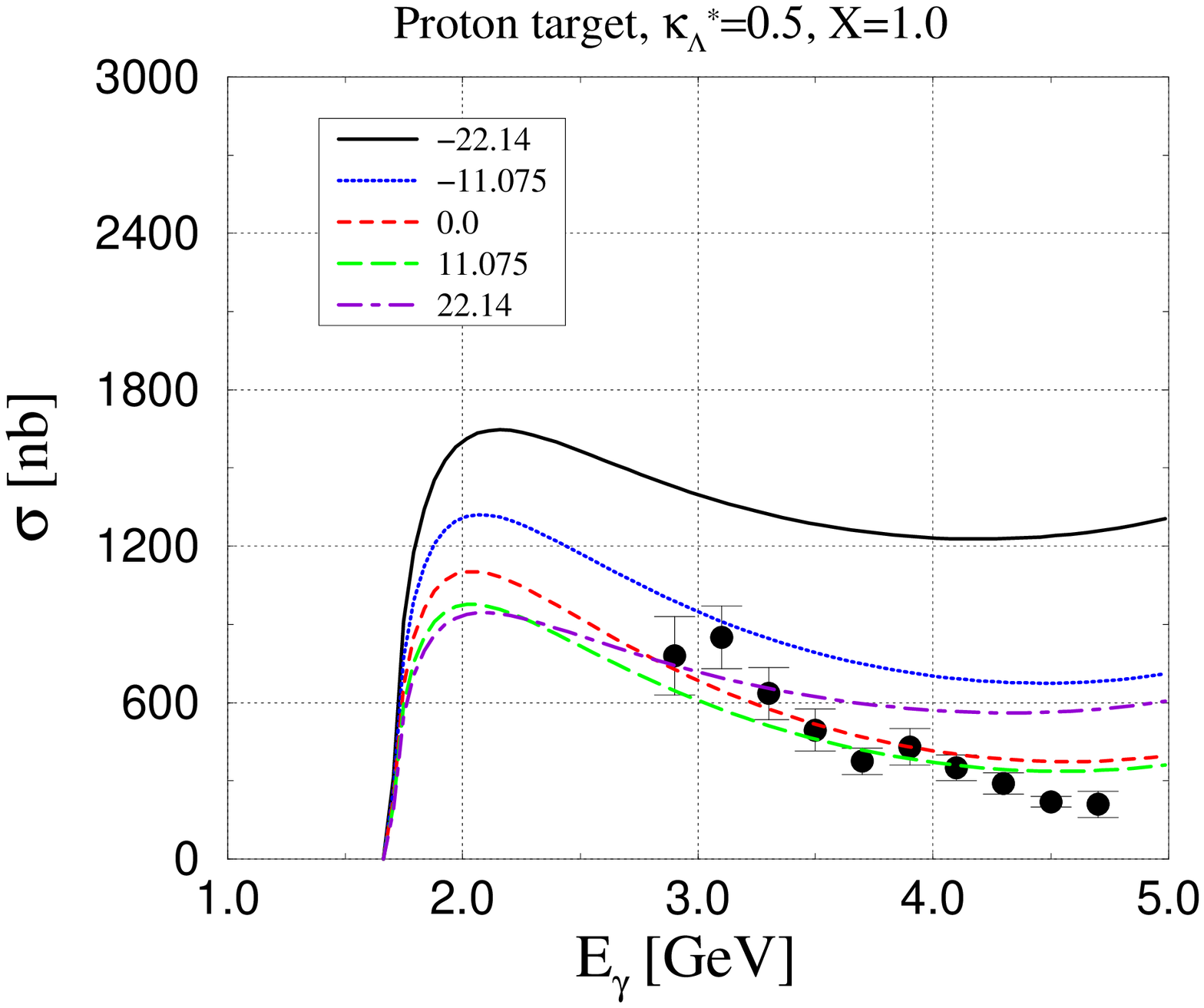}}
\resizebox{5cm}{5cm}{\includegraphics{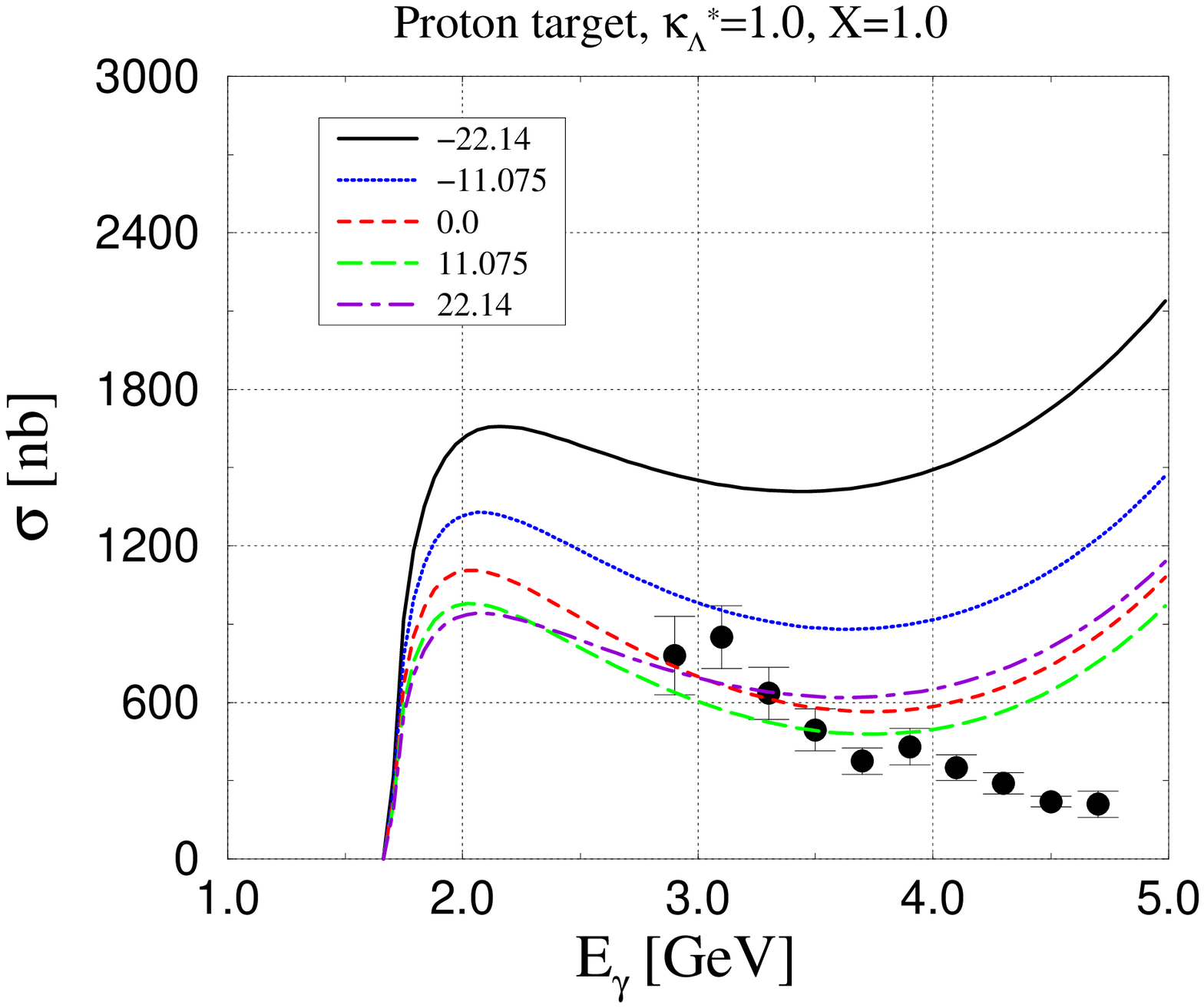}}
\end{tabular}
\caption{The total cross sections for the proton target by changing the
model parameters which are $\kappa_{\Lambda^*}$, $X$ and $g_{K^*N\Lambda^*}$.}
\label{fig1}
\end{figure}

\begin{figure}[tbh]
\begin{tabular}{ccc}
\resizebox{5cm}{5cm}{\includegraphics{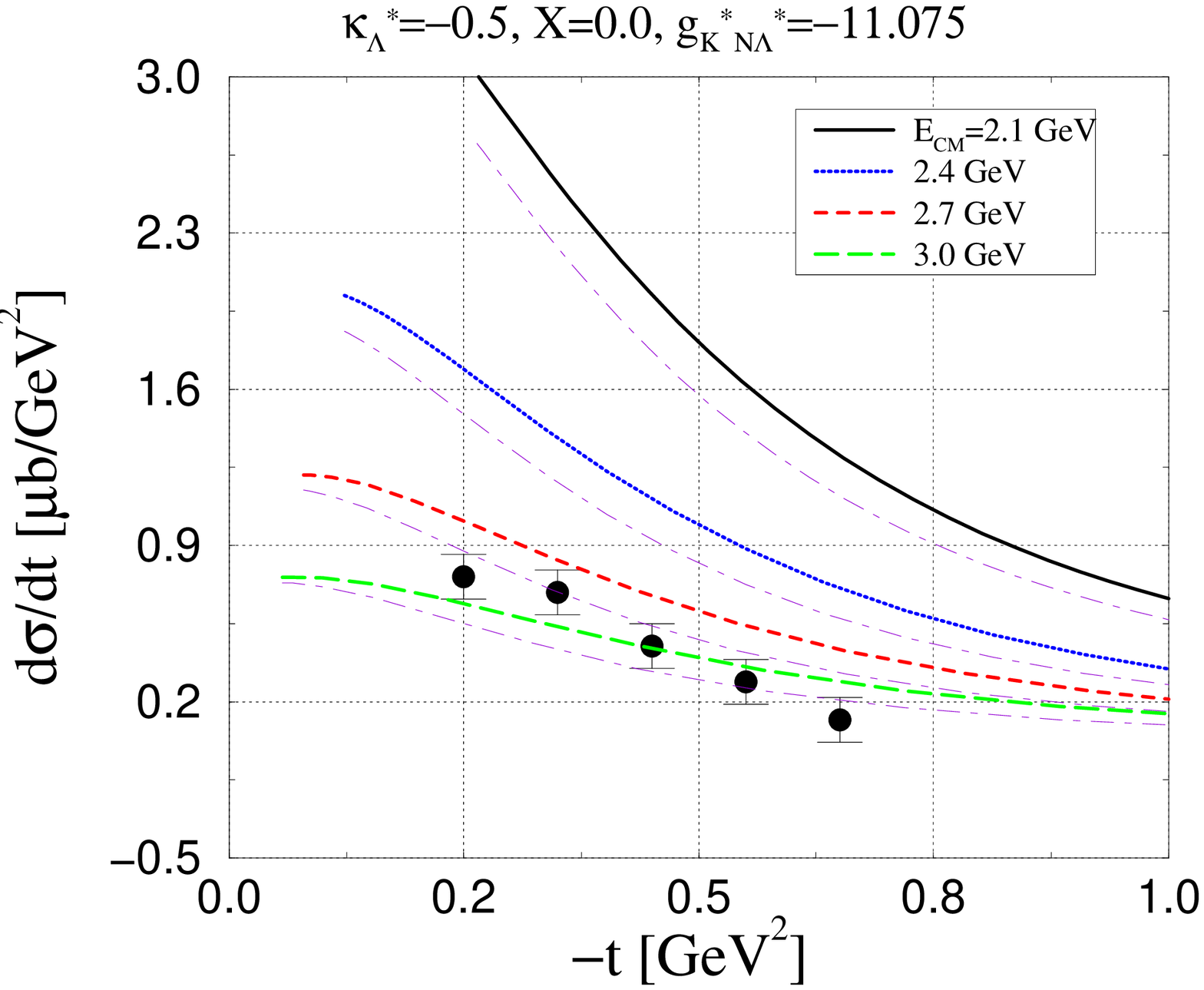}}
\resizebox{5cm}{5cm}{\includegraphics{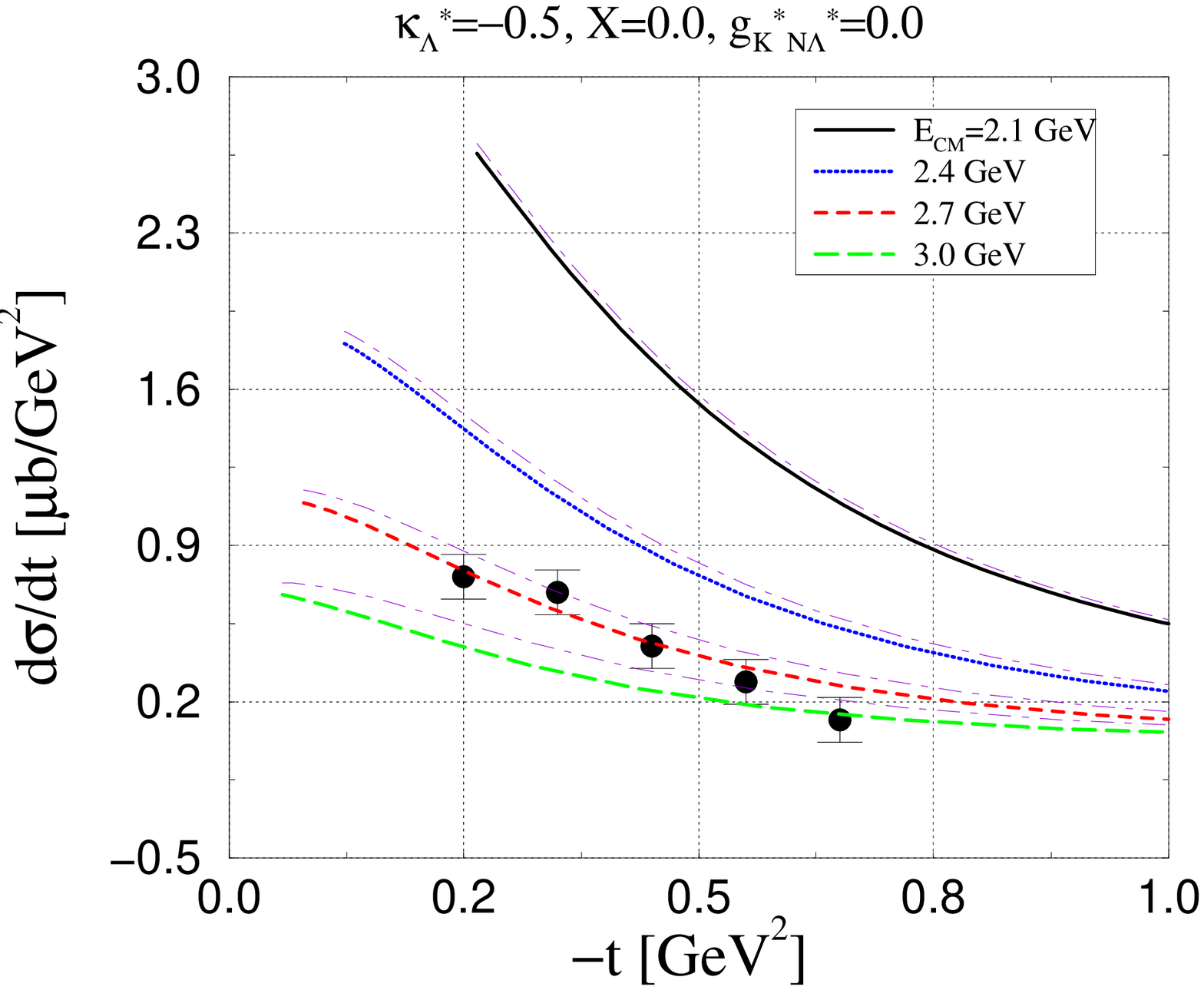}}
\resizebox{5cm}{5cm}{\includegraphics{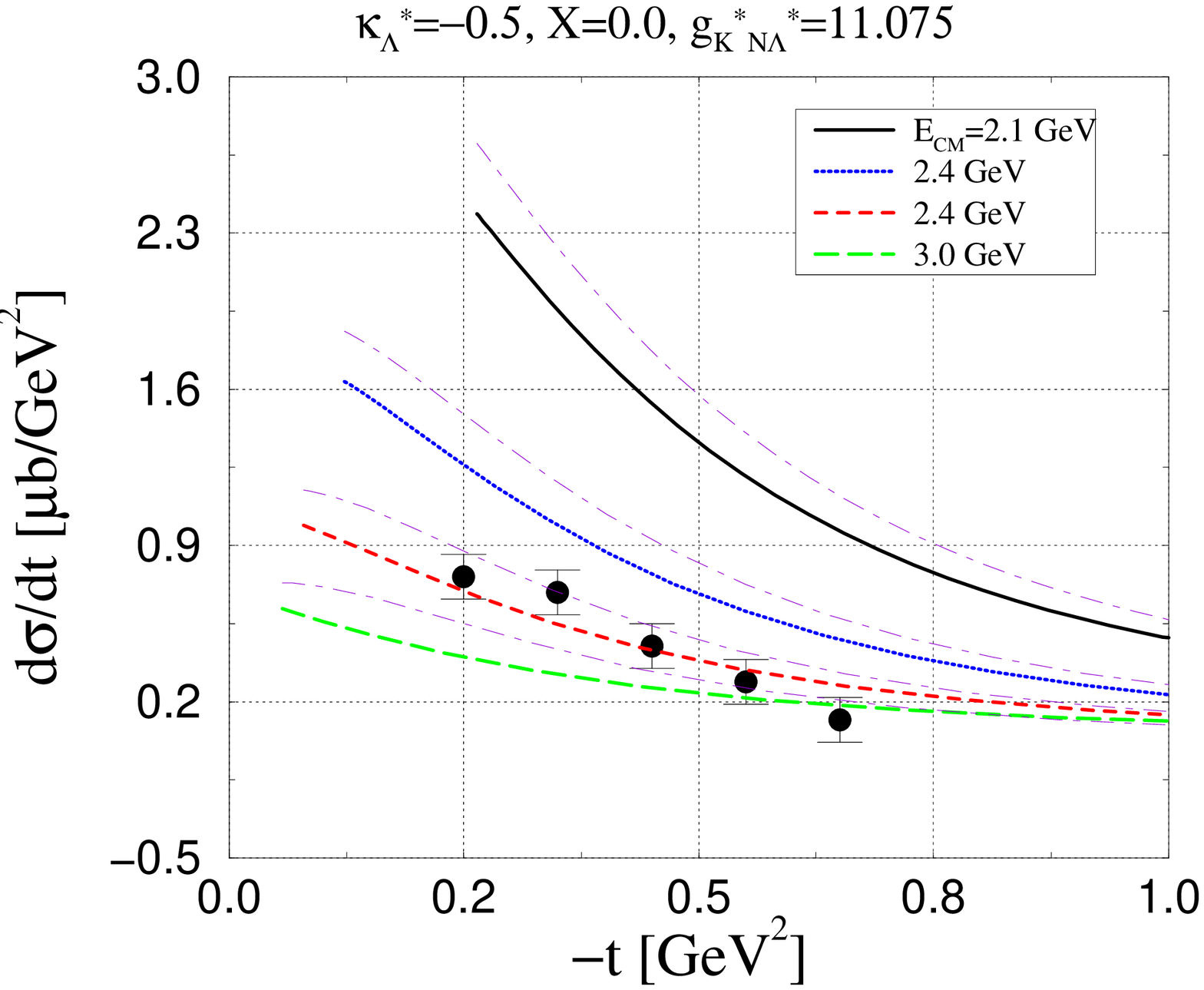}}
\end{tabular}
\caption{The momentum transfer dependence for the proton target
  changing the model parameters which are $\kappa_{\Lambda^*}$, $X$
  and $g_{K^*N\Lambda^*}$. Dot-dashed lines indicated the case without
  the model parameters.}
\label{fig2}
\end{figure} 

\begin{figure}[tbh]
\begin{center}
\resizebox{5cm}{5cm}{\includegraphics{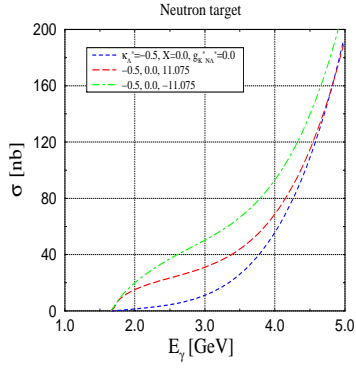}}
\caption{The total cross sections for the neutron
  target changing the model parameters. }
\end{center}
\label{fig3}
\end{figure} 

\begin{figure}[tbh]
\begin{tabular}{ccc}
\resizebox{5cm}{5cm}{\includegraphics{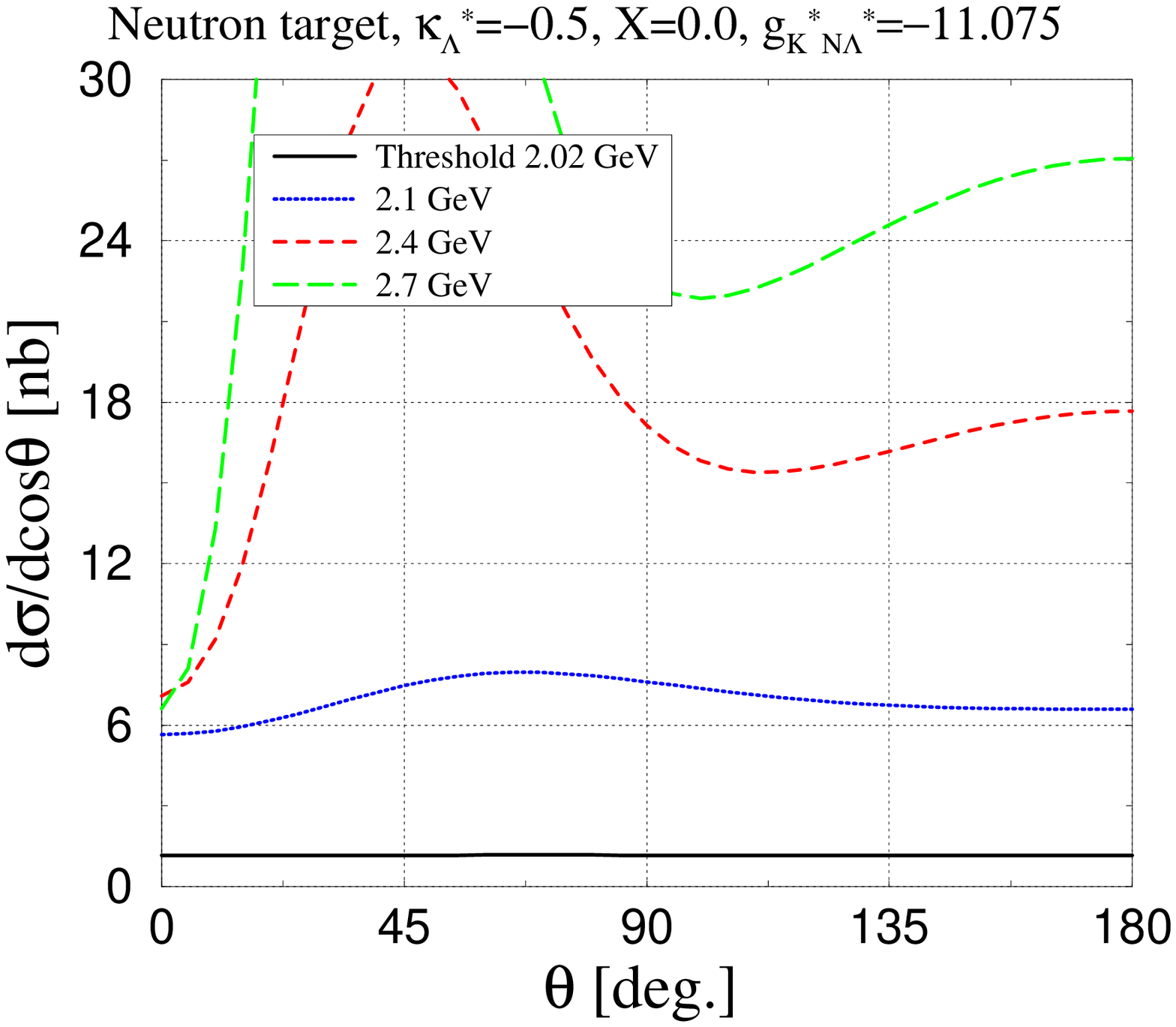}}
\resizebox{5cm}{5cm}{\includegraphics{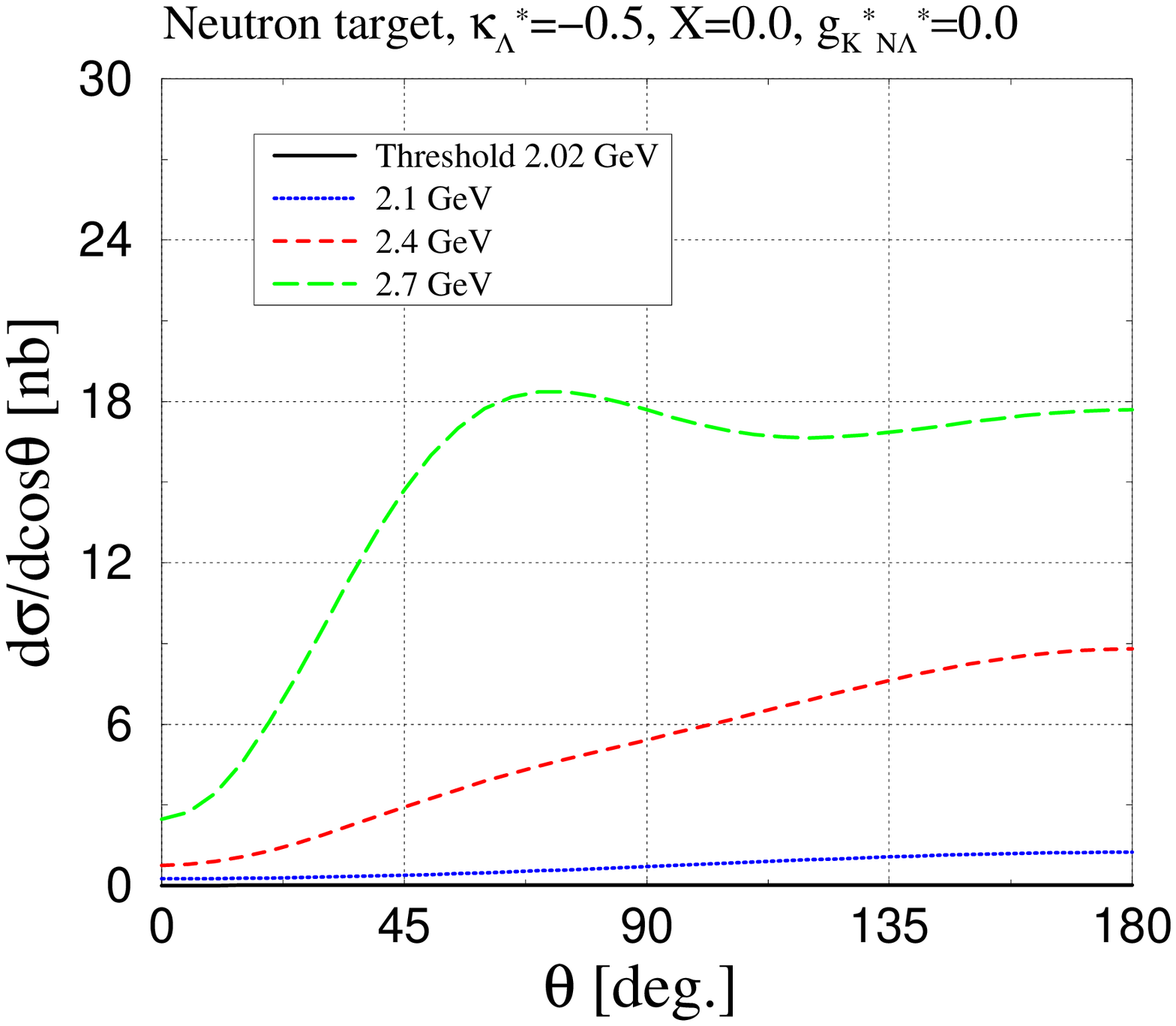}}
\resizebox{5cm}{5cm}{\includegraphics{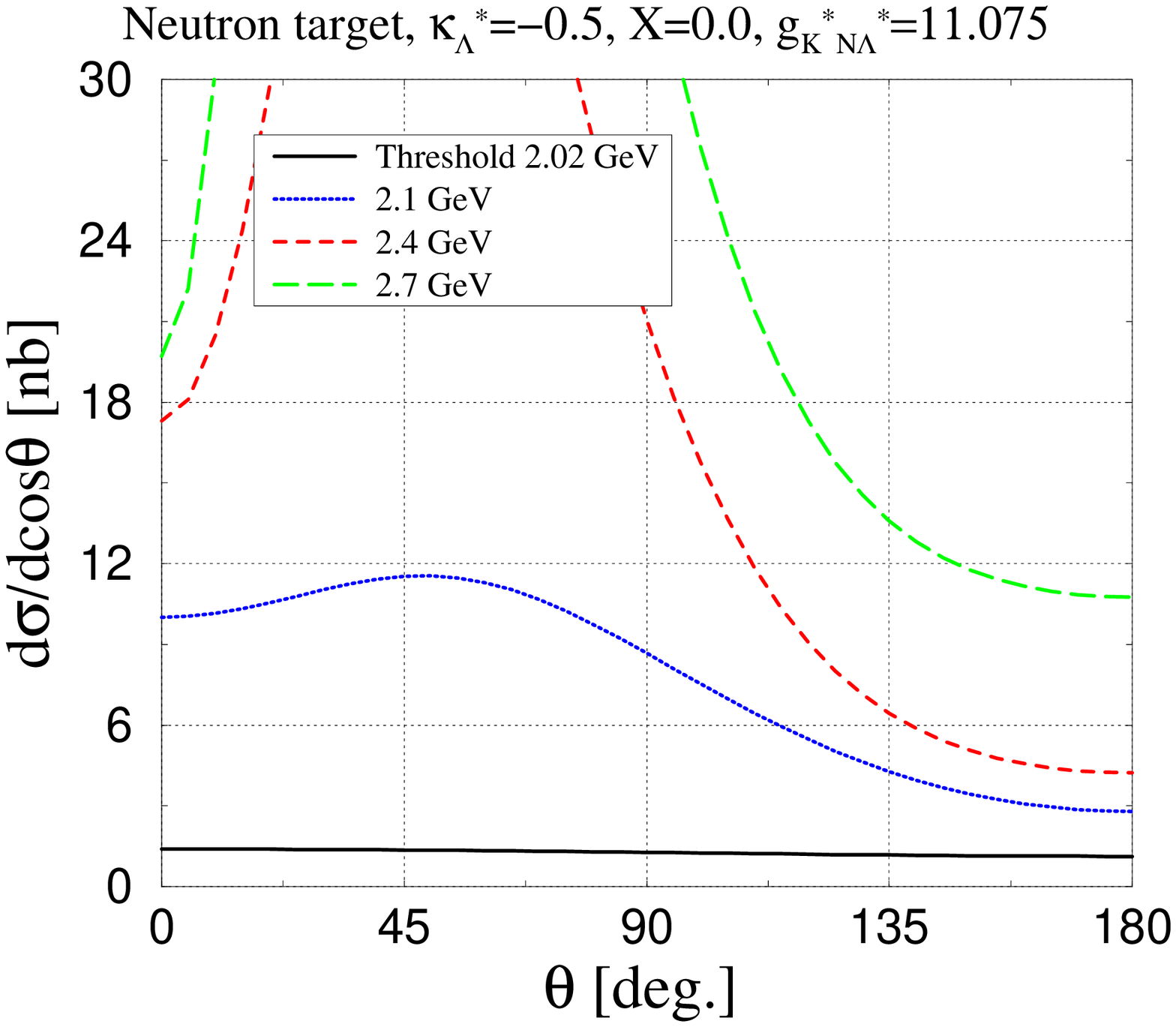}}
\end{tabular}
\caption{The angular distributions for the neutron targetwith the parameter sets.}
\label{fig4}
\end{figure} 
\end{document}